\documentclass[preprint,prd,aps,showpacs,showkeys,nofootinbib]{revtex4}
\usepackage{graphicx}
\usepackage{dcolumn}
\usepackage{bm}
\usepackage{color}
\usepackage[dvipsnames]{xcolor}
\usepackage{amssymb}

\definecolor{light-gray}{gray}{0.78}
\definecolor{mid-gray}{gray}{0.55}
\definecolor{dark-gray}{gray}{0.32}

\begin{document}

\title{$B^{0}-\bar{B^{0}}$ mixing in the $U(1)_X$SSM}
\author{Xing-Yu Han$^{1,2,3}$, Shu-Min Zhao$^{1,2,3}$\footnote{zhaosm@hbu.edu.cn}, Xi Wang$^{1,2,3}$, Yi-Tong Wang$^{1,2,3}$, Tong-Tong Wang$^{1,2,3}$, Xin-Xin Long$^{1,2,3}$, Xing-Xing Dong$^{1,2,3}$, Tai-Fu Feng$^{1,2,4}$}

\affiliation{$^1$ Department of Physics, Hebei University, Baoding 071002, China}
\affiliation{$^2$ Key Laboratory of High-precision Computation and Application of Quantum Field Theory of Hebei Province, Baoding 071002, China}
\affiliation{$^3$ Research Center for Computational Physics of Hebei Province, Baoding 071002, China}
\affiliation{$^4$ Department of Physics, Chongqing University, Chongqing 401331, China}
\date{\today}

\begin{abstract}
$U(1)_X$SSM is a non-universal Abelian extension of the Minimal Supersymmetric Standard Model (MSSM) and its local gauge group is
extended to $SU(3)_C\times SU(2)_L \times U(1)_Y\times U(1)_X$. Based on the latest data of neutral meson mixing and experimental limitations, we investigate the process of $B^{0}-\bar{B^{0}}$ mixing in $U(1)_X$SSM. Using the effective Hamiltonian method, the Wilson coefficients and mass difference $\triangle m_{B}$ are derived. The abundant numerical results verify that $~v_S,~M^2_D,~\lambda_C,~{\mu},~M_2,~\tan{\beta},~g_{YX},~M_1$ and $~\lambda_H$ are sensitive parameters to the process of $B^{0}-\bar{B^{0}}$ mixing. With further measurement in the experiment, the parameter space of the  $U(1)_X$SSM will be further constrained during the mixing process of $B^{0}-\bar{B^{0}}$.
\end{abstract}


\maketitle

\section{Introduction}

The Standard Model (SM) theory of particle physics was gradually established and developed by Glashow, Weinberg, Salam and others to further study the properties and interactions of particles \cite{b0,b1,b2,b3}. It unifies the three basic interactions of strong, weak, and electromagnetic, and has achieved great success. However the SM still cannot explain some physical phenomena, such as the absence of gravity, the problem of gauge hierarchy, dark matter and dark energy, etc. Based on the new symmetry that combines the spatiotemporal symmetry and internal symmetry, physicists extend the SM to produce the Minimal Supersymmetric Standard Model (MSSM) \cite{n0,n1,n2}. Although the MSSM has successfully solved the problems of gauge hierarchy and dark matter, it has not yet solved the problem of neutrino mass and the $\mu$ problem. Because the neutrino experiment results show that neutrino has a small mass, and different generations of neutrino can mix with each other, physicists extend the MSSM using the local gauge group $U(1)_X$ to obtain the $U(1)_X$SSM \cite{sm2}. This new physical model can explain the results of neutrino oscillation experiment when light neutrinos obtain tiny masses by the seesaw mechanism, and relieve $\mu$ problem and the little hierarchy problem in MSSM by the right-handed neutrinos, sneutrinos and additional Higgs singlets.

The flavor changing neutral current (FCNC) process of $b \to s\gamma$, ${K^0} - {{\bar
K}^0}$ and ${B^0} - {{\bar B}^0}$ mixing have played a significant role in particle physics \cite{b4,b5,b6}. They are highly suppressed in the SM and therefore extremely useful for exploring new physical (NP) beyond SM. In 2001, CP violation of the neutral B meson system was observed, and B-system decays have superiority over the K-system to offer a direct test of the CP violation of SM and is free of corrections from strong interactions \cite{n12,n15}. The latest average experimental result of mass difference is \cite{pdg2022}
\begin{eqnarray}
&&\Delta m_{B}^{Exp} =  (3.334\pm0.013)\times 10^{-13}~{\rm{GeV}}.
\end{eqnarray}
The ${B^0} - {{\bar B}^0}$ mixing process has also been calculated in models such as SM and MSSM \cite{m1,m3,m5,m6,m7,m8}. Recently, people have also studied the contribution of ${B^0} - {{\bar B}^0}$ mixing to NP \cite{m10}. The ${B^0} - {{\bar B}^0}$ mixing in a supersymmetric extension of the standard model where baryon and lepton numbers are local gauge symmetries(BLMSSM) shows that the parameters $\lambda_{1,3}$, $m_{\tilde{D_{5}}}$ and $\mu_{X}$ are sensitive to the process of ${B^0} - {{\bar B}^0}$ mixing \cite{m0}. $U(1)_X$SSM can provide new FCNC at loop level in the $B^{0}-\bar{B^{0}}$ mixing, we will calculate  $B^{0}-\bar{B^{0}}$ mixing through the effective Hamiltonian method in this model.

In the following, we mainly introduce the
$U(1)_X$SSM including its superpotential, the general soft breaking terms, the mass matrices and couplings. In Sec.III, we give the analytical
formulae of the $B^{0}-\bar{B^{0}}$ mixing in $U(1)_X$SSM. The corresponding parameters and numerical analysis are shown in Sec.IV.
The last section presents our conclusions. Finally, the Appendix introduces some formulae that we need for this work.

\section{The relevant content of $U(1)_X$SSM}
We extend the MSSM using the local gauge group $U(1)_X$ to obtain the $U(1)_X$SSM with the local gauge group $SU(3)_C\times SU(2)_L \times U(1)_Y\times U(1)_X$.
$U(1)_X$SSM has new superfields beyond  MSSM, including three Higgs singlets $\hat{\eta},~\hat{\bar{\eta}},~\hat{S}$, and right-handed neutrinos $\hat{\nu}_i$.
In order to obtain Higgs boson mass of 125.25 GeV, it is necessary to consider loop correction \cite{n16,n17,n18}.
The particle content and charge distribution of $U(1)_X$SSM can be found in previous studies \cite{n19}.

The superpotential in $U(1)_X$SSM is written as
\begin{eqnarray}
&&W=l_W\hat{S}+\mu\hat{H}_u\hat{H}_d+M_S\hat{S}\hat{S}-Y_d\hat{d}\hat{q}\hat{H}_d-Y_e\hat{e}\hat{l}\hat{H}_d+\lambda_H\hat{S}\hat{H}_u\hat{H}_d
\nonumber\\&&\hspace{0.6cm}+\lambda_C\hat{S}\hat{\eta}\hat{\bar{\eta}}+\frac{\kappa}{3}\hat{S}\hat{S}\hat{S}+Y_u\hat{u}\hat{q}\hat{H}_u+Y_X\hat{\nu}\hat{\bar{\eta}}\hat{\nu}
+Y_\nu\hat{\nu}\hat{l}\hat{H}_u.
\end{eqnarray}
We present the explicit forms of two Higgs doublets and three Higgs singlets here
\begin{eqnarray}
&&\hspace{1cm}H_{u}=\left(\begin{array}{c}H_{u}^+\\{1\over\sqrt{2}}\Big(v_{u}+H_{u}^0+iP_{u}^0\Big)\end{array}\right),
~~~~~~
H_{d}=\left(\begin{array}{c}{1\over\sqrt{2}}\Big(v_{d}+H_{d}^0+iP_{d}^0\Big)\\H_{d}^-\end{array}\right),
\nonumber\\&&\eta={1\over\sqrt{2}}\Big(v_{\eta}+\phi_{\eta}^0+iP_{\eta}^0\Big),~~~
\bar{\eta}={1\over\sqrt{2}}\Big(v_{\bar{\eta}}+\phi_{\bar{\eta}}^0+iP_{\bar{\eta}}^0\Big),~~
S={1\over\sqrt{2}}\Big(v_{S}+\phi_{S}^0+iP_{S}^0\Big).
\end{eqnarray}
$v_u,~v_d,~v_\eta$, $v_{\bar\eta}$ and $v_S$ are the corresponding vacuum expectation values(VEVs) of the Higgs superfields $H_u$, $H_d$, $\eta$, $\bar{\eta}$ and $S$.

The soft SUSY breaking terms are generally given as
\begin{eqnarray}
&&\mathcal{L}_{soft}=\mathcal{L}_{soft}^{MSSM}-B_SS^2-L_SS-\frac{T_\kappa}{3}S^3-T_{\lambda_C}S\eta\bar{\eta}
+\epsilon_{ij}T_{\lambda_H}SH_d^iH_u^j\nonumber\\&&\hspace{1cm}
-T_X^{IJ}\bar{\eta}\tilde{\nu}_R^{*I}\tilde{\nu}_R^{*J}
+\epsilon_{ij}T^{IJ}_{\nu}H_u^i\tilde{\nu}_R^{I*}\tilde{l}_j^J
-m_{\eta}^2|\eta|^2-m_{\bar{\eta}}^2|\bar{\eta}|^2-m_S^2S^2\nonumber\\&&\hspace{1cm}
-(m_{\tilde{\nu}_R}^2)^{IJ}\tilde{\nu}_R^{I*}\tilde{\nu}_R^{J}
-\frac{1}{2}\Big(M_S\lambda^2_{\tilde{X}}+2M_{BB^\prime}\lambda_{\tilde{B}}\lambda_{\tilde{X}}\Big)+h.c~.
\end{eqnarray}
Here is the covariant derivatives of $U(1)_X$SSM
\begin{eqnarray}
&&D_\mu=\partial_\mu-i\left(\begin{array}{cc}Y^Y,&Y^X\end{array}\right)
\left(\begin{array}{cc}g_{1},&g_{{YX}}\\0,&g_{{X}}\end{array}\right)
\left(\begin{array}{c}A_{\mu}^{Y} \\ A_{\mu}^{X}\end{array}\right)\;.
\end{eqnarray}

The mass matrix for neutralino in the basis $(\lambda_{\tilde{B}}, \tilde{W}^0, \tilde{H}_d^0, \tilde{H}_u^0,
\lambda_{\tilde{X}}, \tilde{\eta}, \tilde{\bar{\eta}}, \tilde{s}) $ is

\begin{equation}
m_{\tilde{\chi}^0} = \left(
\begin{array}{cccccccc}
M_1 &0 &-\frac{g_1}{2}v_d &\frac{g_1}{2}v_u &{M}_{B B'} &0  &0  &0\\
0 &M_2 &\frac{1}{2} g_2 v_d  &-\frac{1}{2} g_2 v_u  &0 &0 &0 &0\\
-\frac{g_1}{2}v_d &\frac{1}{2} g_2 v_d  &0
&m_{\tilde{H}_d^0\tilde{H}_u^0} &m_{\tilde{H}_d^0\lambda_{\tilde{X}}} &0 &0 & - \frac{{\lambda}_{H} v_u}{\sqrt{2}}\\
\frac{g_1}{2}v_u &-\frac{1}{2} g_2 v_u  &m_{\tilde{H}_d^0\tilde{H}_u^0} &0 &m_{\tilde{H}_u^0\lambda_{\tilde{X}}} &0 &0 &- \frac{{\lambda}_{H} v_d}{\sqrt{2}}\\
{M}_{B B'} &0 &m_{\tilde{H}_d^0\lambda_{\tilde{X}}} &m_{\tilde{H}_u^0\lambda_{\tilde{X}}} &{M}_{BL} &- g_{X} v_{\eta}  &g_{X} v_{\bar{\eta}}  &0\\
0  &0 &0 &0 &- g_{X} v_{\eta}  &0 &\frac{1}{\sqrt{2}} {\lambda}_{C} v_S  &\frac{1}{\sqrt{2}} {\lambda}_{C} v_{\bar{\eta}} \\
0  &0 &0 &0 &g_{X} v_{\bar{\eta}}  &\frac{1}{\sqrt{2}} {\lambda}_{C} v_S  &0 &\frac{1}{\sqrt{2}} {\lambda}_{C} v_{\eta} \\
0 &0 & - \frac{{\lambda}_{H} v_u}{\sqrt{2}} &- \frac{{\lambda}_{H} v_d}{\sqrt{2}} &0 &\frac{1}{\sqrt{2}} {\lambda}_{C} v_{\bar{\eta}}
 &\frac{1}{\sqrt{2}} {\lambda}_{C} v_{\eta}  &m_{\tilde{s}\tilde{s}}\end{array}
\right),\label{neutralino}
 \end{equation}

\begin{eqnarray}
&& m_{\tilde{H}_d^0\tilde{H}_u^0} = - \frac{1}{\sqrt{2}} {\lambda}_{H} v_S  - \mu ,~~~~~~~
m_{\tilde{H}_d^0\lambda_{\tilde{X}}} = -\frac{1}{2} (g_{Y X} + g_{X})v_d, \nonumber\\&&
m_{\tilde{H}_u^0\lambda_{\tilde{X}}} = \frac{1}{2} (g_{Y X} + g_{X})v_u
 ,~~~~~~~~~~~~
m_{\tilde{s}\tilde{s}} = 2 M_S  + \sqrt{2} \kappa v_S.\label{neutralino1}
\end{eqnarray}
This matrix is diagonalized by $Z^N$,
\begin{equation}
Z^{N*} m_{\tilde{\chi}^0} Z^{N{\dagger}} = m^{diag}_{\tilde{\chi}^0}.
\end{equation}
One can find other mass matrixes in the Appendix.

In addition, there are some couplings that need to be used later:
\begin{eqnarray}
&&\mathcal{L}_{\chi^{-}\tilde{u}d}=\bar{\chi}_i^{-}
\Big((V_{i2}^{*}Y_{u,j}Z^U_{k,3+j}-g_2V_{i1}^{*}Z_{kj}^U
)P_L+Y_{d,j}^{*}Z_{k,j}^UU_{i2}P_R\Big)d_{j}\tilde{u}_{k}^*,
\\&&\mathcal{L}_{\chi^0d\tilde{D}}=-\frac{1}{6}\bar{\chi}^0_i\Big\{\Big[\sqrt{2}( g_1 N_{i1}^{*}
-3 g_2 N_{i2}^{*}    + g_{Y X} N_{i5}^{*}) Z_{k,j}^D +6 N_{i3}^{*} Y_{d,j} Z_{k,3+j}^D   \Big]P_L
\nonumber\\&&\hspace{1.2cm}+\Big[6 Y_{d,j}^{*}  Z_{k,j}^D  N_{i3}
+ \sqrt{2} Z_{k,3+j}^D [2 g_1 N_{i1}
 + (2 g_{Y X}  + 3 g_{X})N_{i5}]\Big]P_R\Big\}d_j\tilde{d}^*_k.\label{LX01}
\end{eqnarray}
To save space in the text, the remaining vertexes can be found in the Appendix.

\section{Analytical formula}

The effective Hamiltonian method, as the primary technique used in the computation of neutral meson mixing \cite{m11,m12,m13}, involves expressing the effective Hamiltonian of the system as a product of the Wilson coefficient and the effective operator, which is obtained using the operator product expansion (OPE). OPE separates the short distance contributions $C_i (\mu)$ and long distance contributions $Q_i (\mu)$, with the former being expressed by the Wilson coefficient through perturbation methods and the latter requiring non-perturbation techniques like lattice QCD, QCD sum rule, $\frac{1}{N}$ expansion, etc. The energy scale then evolves from $\mu=O (m_W)$ to the hadron energy scale. Notably, the amplitude is independent of the chosen energy scale, allowing for the offsetting of the energy scale dependence of the Wilson coefficient and operator.

The general form of the effective Hamiltonian for $B^{0}-\bar{B^{0}}$ mixture under the weak energy scale can be expressed as \cite{n20}
\begin{eqnarray}
{H_{eff}} =\frac{1}{4} \frac{{{\rm{G}}_{\rm{F}}^2}}{{{\pi ^2}}}m_W^2\sum_{\alpha=1}^{8}C_{\alpha}{{{\cal O}_{\alpha}}},
\label{Heff}
\end{eqnarray}
Where ${G_F}$ denotes the Fermi constant,  ${C_\alpha } $ are the corresponding Wilson coefficients,
${{{{\cal O}}}_\alpha }$ are the effective operators,
\begin{eqnarray}
&&{\mathcal{O}_1} = \bar d{\gamma _\mu }{P_L}b\bar d{\gamma ^\mu }{P_L}b,
\nonumber\\
&&{\mathcal{O}_2} = \bar d{\gamma _\mu }{P_L}b\bar d{\gamma ^\mu }{P_R}b,
\nonumber\\
&&{\mathcal{O}_3} = \bar d{P_L}b\bar d{P_R}b,
\nonumber\\
&&{\mathcal{O}_4} = \bar d{P_L}b\bar d{P_L}b,
\nonumber\\
&&{\mathcal{O}_5} = \bar d{\sigma _{\mu \nu }}{P_L}b\bar d{\sigma ^{\mu \nu }}{P_L}b,
\nonumber\\
&&{\mathcal{O}_6} = \bar d{\gamma _\mu }{P_R}b\bar d{\gamma ^\mu }{P_R}b,
\nonumber\\
&&{\mathcal{O}_7} = \bar d{P_R}b\bar d{P_R}b,
\nonumber\\
&&{\mathcal{O}_8} = \bar d{\sigma _{\mu \nu }}{P_R}b\bar d{\sigma ^{\mu \nu }}{P_R}b.
\end{eqnarray}
Here, ${P_{R,L}} = \left( {1 \pm {\gamma _5}} \right)/2$ denote the chiral projectors, ${\sigma _{\mu \nu }} = \left[ {{\gamma _\mu }, {\gamma _\nu }} \right]/2$.

\begin{figure}[ht]
\setlength{\unitlength}{5.0mm}
\centering
\includegraphics[width=5.0in]{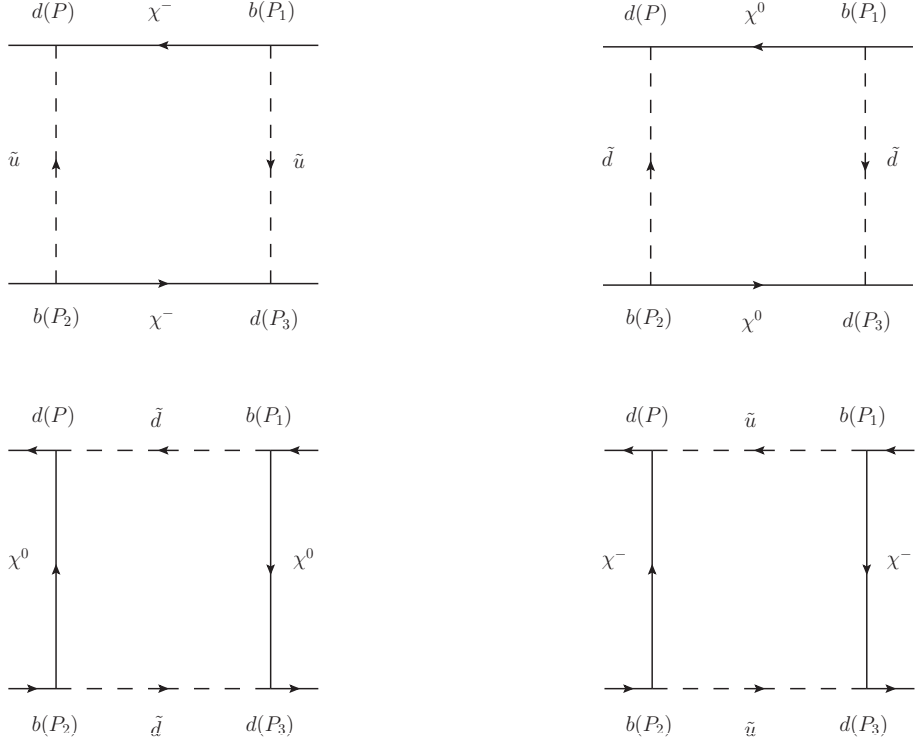}
\caption{The box diagrams contributing to ${B^0} - {{\bar B}^0}$ mixing in the $U(1)_X$SSM.}\label{N1}
\end{figure}

The box diagrams contributing to ${B^0} - {{\bar B}^0}$ mixing in the $U(1)_X$SSM are shown in Fig.\ref{N1}. Among then, the diagrams including the particles $\tilde{\chi}^0$ and $\chi^{-}$ should make a Fierz rearrangement.

The Wilson coefficients can be expressed as
\begin{eqnarray}
&&\hspace{-0.5cm}C_1=\frac{2 \pi ^2 }{G^2_F m^2_W}\Big( {\sum_{b,n=1}^8\sum_{k,m=1}^6B_{24}}(m_{\tilde{\chi}^0_b}^2,m_{\tilde{\chi}^0_n}^2,m_{\tilde{d}_k},m_{\tilde{d}_m})A^{d\tilde{u}\chi^{-}}_RB^{\chi^{-}\tilde{u}d}_LC^{\chi^{-}\tilde{u}d}_LD^{d\tilde{u}\chi^{-}}_R
\nonumber\\&&\hspace{0.5cm}+\sum_{b,n=1}^2\sum_{k,m=1}^6{B_{24}}(m_{\chi^{-}_b}^2,m_{\chi^{-}_n}^2,m_{\tilde{u}_k},m_{\tilde{u}_m})A^{d\tilde{d}\tilde{\chi}^0}_RB^{\tilde{\chi}^0\tilde{d}d}_LC^{\tilde{\chi}^0\tilde{d}d}_LD^{d\tilde{d}\tilde{\chi}^0}_R\Big),
\nonumber\\
&&\hspace{-0.5cm}C_2=\frac{-2 \pi ^2 }{G^2_F m^2_W}\Big( \sum_{b,n=1}^8\sum_{k,m=1}^6m_{\tilde{\chi}^0_b}^2m_{\tilde{\chi}^0_n}^2{B_{04}}(m_{\tilde{\chi}^0_b}^2,m_{\tilde{\chi}^0_n}^2,m_{\tilde{d}_k},m_{\tilde{d}_m})A^{d\tilde{u}\chi^{-}}_LB^{\chi^{-}\tilde{u}d}_LC^{\chi^{-}\tilde{u}d}_RD^{d\tilde{u}\chi^{-}}_R
\nonumber\\&&\hspace{0.5cm}+\sum_{b,n=1}^2\sum_{k,m=1}^6m_{\chi^{-}_b}^2m_{\chi^{-}_n}^2{B_{04}}(m_{\chi^{-}_b}^2,m_{\chi^{-}_n}^2,m_{\tilde{u}_k},m_{\tilde{u}_m})A^{d\tilde{d}\tilde{\chi}^0}_LB^{\tilde{\chi}^0\tilde{d}d}_LC^{\tilde{\chi}^0\tilde{d}d}_RD^{d\tilde{d}\tilde{\chi}^0}_R
\nonumber\\&&\hspace{0.5cm}+ \sum_{b,n=1}^8\sum_{k,m=1}^6m_{\tilde{\chi}^0_b}^2m_{\tilde{\chi}^0_n}^2{B_{04}}(m_{\tilde{\chi}^0_b}^2,m_{\tilde{\chi}^0_n}^2,m_{\tilde{d}_k},m_{\tilde{d}_m})A^{d\tilde{u}\chi^{-}}_LB^{\chi^{-}\tilde{u}d}_RC^{\chi^{-}\tilde{u}d}_LD^{d\tilde{u}\chi^{-}}_R
\nonumber\\&&\hspace{0.5cm}+\sum_{b,n=1}^2\sum_{k,m=1}^6m_{\chi^{-}_b}^2m_{\chi^{-}_n}^2{B_{04}}(m_{\chi^{-}_b}^2,m_{\chi^{-}_n}^2,m_{\tilde{u}_k},m_{\tilde{u}_m})A^{d\tilde{d}\tilde{\chi}^0}_LB^{\tilde{\chi}^0\tilde{d}d}_RC^{\tilde{\chi}^0\tilde{d}d}_LD^{d\tilde{d}\tilde{\chi}^0}_L\Big),
\nonumber\\
&&\hspace{-0.5cm}C_3=\frac{-2 \pi ^2 }{G^2_F m^2_W}\Big( \sum_{b,n=1}^8\sum_{k,m=1}^6{B_{24}}(m_{\tilde{\chi}^0_b}^2,m_{\tilde{\chi}^0_n}^2,m_{\tilde{d}_k},m_{\tilde{d}_m})A^{d\tilde{u}\chi^{-}}_LB^{\chi^{-}\tilde{u}d}_RC^{\chi^{-}\tilde{u}d}_LD^{d\tilde{u}\chi^{-}}_R
\nonumber\\&&\hspace{0.5cm}+\sum_{b,n=1}^2\sum_{k,m=1}^6{B_{24}}(m_{\chi^{-}_b}^2,m_{\chi^{-}_n}^2,m_{\tilde{u}_k},m_{\tilde{u}_m})A^{d\tilde{d}\tilde{\chi}^0}_LB^{\tilde{\chi}^0\tilde{d}d}_RC^{\tilde{\chi}^0\tilde{d}d}_LD^{d\tilde{d}\tilde{\chi}^0}_L
\nonumber\\&&\hspace{0.5cm}+\sum_{b,n=1}^8\sum_{k,m=1}^6{B_{24}}(m_{\tilde{\chi}^0_b}^2,m_{\tilde{\chi}^0_n}^2,m_{\tilde{d}_k},m_{\tilde{d}_m})A^{d\tilde{u}\chi^{-}}_RB^{\chi^{-}\tilde{u}d}_RC^{\chi^{-}\tilde{u}d}_LD^{d\tilde{u}\chi^{-}}_L
\nonumber\\&&\hspace{0.5cm}+\sum_{b,n=1}^2\sum_{k,m=1}^6{B_{24}}(m_{\chi^{-}_b}^2,m_{\chi^{-}_n}^2,m_{\tilde{u}_k},m_{\tilde{u}_m})A^{d\tilde{d}\tilde{\chi}^0}_RB^{\tilde{\chi}^0\tilde{d}d}_RC^{\tilde{\chi}^0\tilde{d}d}_LD^{d\tilde{d}\tilde{\chi}^0}_L\Big),
\nonumber\\
&&\hspace{-0.5cm}C_4=\frac{-4 \pi ^2 }{G^2_F m^2_W}\Big( \sum_{b,n=1}^8\sum_{k,m=1}^6m_{\tilde{\chi}^0_b}^2m_{\tilde{\chi}^0_n}^2{B_{04}}(m_{\tilde{\chi}^0_b}^2,m_{\tilde{\chi}^0_n}^2,m_{\tilde{d}_k},m_{\tilde{d}_m})A^{d\tilde{u}\chi^{-}}_LB^{\chi^{-}\tilde{u}d}_LC^{\chi^{-}\tilde{u}d}_LD^{d\tilde{u}\chi^{-}}_L
\nonumber\\&&\hspace{0.5cm}+\sum_{b,n=1}^2\sum_{k,m=1}^6m_{\chi^{-}_b}^2m_{\chi^{-}_n}^2{B_{04}}(m_{\chi^{-}_b}^2,m_{\chi^{-}_n}^2,m_{\tilde{u}_k},m_{\tilde{u}_m})A^{d\tilde{d}\tilde{\chi}^0}_LB^{\tilde{\chi}^0\tilde{d}d}_LC^{\tilde{\chi}^0\tilde{d}d}_LD^{d\tilde{d}\tilde{\chi}^0}_L\Big),
\nonumber\\
&&\hspace{-0.5cm}C_6=\frac{2 \pi ^2 }{G^2_F m^2_W}\Big( \sum_{b,n=1}^8\sum_{k,m=1}^6{B_{24}}(m_{\tilde{\chi}^0_b}^2,m_{\tilde{\chi}^0_n}^2,m_{\tilde{d}_k},m_{\tilde{d}_m})A^{d\tilde{u}\chi^{-}}_LB^{\chi^{-}\tilde{u}d}_RC^{\chi^{-}\tilde{u}d}_RD^{d\tilde{u}\chi^{-}}_L
\nonumber\\&&\hspace{0.5cm}+\sum_{b,n=1}^2\sum_{k,m=1}^6{B_{24}}(m_{\chi^{-}_b}^2,m_{\chi^{-}_n}^2,m_{\tilde{u}_k},m_{\tilde{u}_m})A^{d\tilde{d}\tilde{\chi}^0}_LB^{\tilde{\chi}^0\tilde{d}d}_RC^{\tilde{\chi}^0\tilde{d}d}_RD^{d\tilde{d}\tilde{\chi}^0}_L\Big),
\nonumber\\
&&\hspace{-0.5cm}C_7=\frac{-4 \pi ^2 }{G^2_F m^2_W}\Big( \sum_{b,n=1}^8\sum_{k,m=1}^6m_{\tilde{\chi}^0_b}^2 m_{\tilde{\chi}^0_n}^2{B_{04}}(m_{\tilde{\chi}^0_b}^2,m_{\tilde{\chi}^0_n}^2,m_{\tilde{d}_k},m_{\tilde{d}_m})A^{d\tilde{u}\chi^{-}}_LB^{\chi^{-}\tilde{u}d}_LC^{\chi^{-}\tilde{u}d}_LD^{d\tilde{u}\chi^{-}}_L
\nonumber\\&&\hspace{0.5cm}+\sum_{b,n=1}^2\sum_{k,m=1}^6m_{\chi^{-}_b}^2m_{\chi^{-}_n}^2{B_{04}}(m_{\chi^{-}_b}^2,m_{\chi^{-}_n}^2,m_{\tilde{u}_k},m_{\tilde{u}_m})A^{d\tilde{d}\tilde{\chi}^0}_RB^{\tilde{\chi}^0\tilde{d}d}_RC^{\tilde{\chi}^0\tilde{d}d}_RD^{d\tilde{d}\tilde{\chi}^0}_R\Big),
\nonumber\\
&&\hspace{-0.5cm}C_5=C_8=0.
\end{eqnarray}
Here, $B_{04}$ and $B_{24}$ are been defined as
\begin{eqnarray}
\mu^{2\epsilon}\int {\frac{{{d^D}P}}{{{{\left( {2\pi } \right)}^D}}}} \frac{1}{{{p^2} - m_1^2}}\frac{1}{{{p^2} - m_2^2}}\frac{1}{{{p^2} - m_3^2}}\frac{1}{{{p^2} - m_4^2}} = \frac{1}{{16{\pi ^2}m_W^4}}{B_{04}}\left( {{x_1},{x_2},{x_3},{x_4}} \right),
\end{eqnarray}
\begin{eqnarray}
\mu^{2\epsilon}\int {\frac{{{d^D}P}}{{{{\left( {2\pi } \right)}^D}}}} \frac{1}{{{p^2} - m_1^2}}\frac{1}{{{p^2} - m_2^2}}\frac{1}{{{p^2} - m_3^2}}\frac{1}{{{p^2} - m_4^2}}{p^2} = \frac{1}{{16{\pi ^2}m_W^2}}{B_{24}}\left( {{x_1},{x_2},{x_3},{x_4}} \right).
\end{eqnarray}
A, B, C, D are coupling constants of the corresponding vertexes mentioned earlier.
$W({\mu _{\rm{b}}},{\mu _W})$ is the leading-order evolution matrix
\begin{eqnarray}
W({\mu _{\rm{b}}},{\mu _W}) = {\left[ {\frac{{{\alpha _S}({m_W})}}{{{\alpha _S}({m_b})}}} \right]^{\frac{{{\gamma ^{(0)}}}}{{2{\beta _0}}}}}.
\end{eqnarray}
$\beta_{0}$ is given by
\begin{eqnarray}
\beta_{0}  = \frac{{11{N_c} - 2{n_f}}}{3},
\end{eqnarray}
where $n_f$ is the number of active flavours, $Nc$
denoting the number of colors and $\gamma^{(0)}$ is the anomalous dimensions matrix (ADM) \cite{n21,n22}.

Through the renormalization-group evolution matrix \cite{n21}, we have
\begin{eqnarray}
{\vec{C}({\mu _{\rm{b}}}) = W({\mu _{\rm{b}}},{\mu _W})\vec{C}({\mu _W})}.
\end{eqnarray}
The mass difference $\triangle m_{B}$ can be expressed as
\begin{eqnarray}
\triangle m_{B}=\frac{1}{4} \frac{{{\rm{G}}_{\rm{F}}^2}}{{{\pi ^2}}}m_W^2\sum_{\alpha=1}^{8}
\frac{{\left| {C_{\alpha}({\mu _{\rm{b}}})\left\langle {{{\bar B}^0}\left|
{{{{\cal O_{\alpha}}}}({\mu _{\rm{b}}})} \right|{B^0}} \right\rangle } \right|}}{{{m_B}}}.
\end{eqnarray}

The hadronic matrix elements can be written as
\[
\begin{array}{l}
 \left\langle {\bar B^0 \left| {{\cal O}_1 } \right|B^0 } \right\rangle  = \frac{2}{3}{B_B}(\mu )f_B^2 m_B^2,
 \nonumber\\
 \left\langle {\bar B^0 \left| {{\cal O}_2 } \right|B^0 } \right\rangle  =  - \frac{1}{{6}}{B_B}(\mu )f_B^2 m_B^2,
 \nonumber\\
 \left\langle {\bar B^0 \left| {{\cal O}_3 } \right|B^0 } \right\rangle  =  - \frac{5}{{12}}{B_B}(\mu )f_B^2 m_B^2,
 \nonumber\\
 \left\langle {\bar B^0 \left| {{\cal O}_4 } \right|B^0 } \right\rangle  = \frac{5}{{12}}{B_B}(\mu )f_B^2 m_B^2,
 \nonumber\\
 \left\langle {\bar B^0 \left| {{\cal O}_5 } \right|B^0 } \right\rangle  = \frac{1}{2}{B_B}(\mu )f_B^2 m_B^2,
 \nonumber\\
 \left\langle {\bar B^0 \left| {{\cal O}_6 } \right|B^0 } \right\rangle  = \frac{2}{3}{B_B}(\mu )f_B^2 m_B^2,
 \nonumber\\
 \left\langle {\bar B^0 \left| {{\cal O}_7 } \right|B^0 } \right\rangle  = \frac{5}{{12}}{B_B}(\mu )f_B^2 m_B^2,
 \nonumber\\
 \left\langle {\bar B^0 \left| {{\cal O}_8 } \right|B^0 } \right\rangle  = \frac{1}{2}{B_B}(\mu )f_B^2 m_B^2.
 \end{array}
\]
Here $f_B$ is the $B$-meson decay constant, $B_{B}$ is  the  bag parameter.
\section{Numerical analysis}
Incorporating certain experimental constraints, this section presents one-dimensional graphs and multidimensional scatter plots. According to the latest LHC data \cite{n23,n24,n25,n26,n27,n28}, we hold  the chargino mass is greater than $1100 ~{\rm GeV}$, the slepton mass is greater than $700~{\rm GeV}$, the squark mass is greater than $1500~{\rm GeV}$, the experimental value of $\tan{\beta}_{\eta}$ should be less than 1.5 and the lightest CP-even Higgs mass $m_{h}$=125.25 GeV \cite{pdg2022}. The parameters we use are as follows:
\begin{eqnarray}
&&~~\kappa=0.1,~~~~~~~M_S =1.2 ~{\rm TeV},~~~~~l_W= 10~{\rm TeV}^2,~\tan{\beta}_{\eta}=0.9,
\nonumber\\&&~M_{BL}=1~{\rm TeV},~~B_{\mu} = 2~{\rm TeV}^2,~~~B_S=1~{\rm TeV}^2,~~~M_{BB'}=0.8~{\rm TeV},
\nonumber\\&&~T_{\lambda_C} =T_{\kappa} =T_{\lambda_H} =1~ {\rm TeV},~~~T_{d12}=T_{d21}=0.25~{\rm TeV}.
\end{eqnarray}

In the following numerical analysis, the parameters needed to study contain:
\begin{eqnarray}
&&~\lambda_C,~M_1,~M_2,~{\mu},~\lambda_H,~M^2_{Djj}=M^2_D,~M^2_{D11},~v_S,~g_{X},~M^2_{Qii}=M^2_Q,
\nonumber\\&&~\tan{\beta},~g_{YX},~T_{uii}=T_u,~T_{dii}=T_d,(i=1,2,3),(j=2,3).
\end{eqnarray}
In addition to the above parameters, the nondiagonal elements of the parameters are defined as zero.

\subsection{The one-dimensional graphs}
In this part, we have fixed the some parameters:
\begin{eqnarray}
&&~\lambda_C=-0.3,~M_2=2.8~{\rm TeV},~{\mu}=1.1~{\rm TeV},~\lambda_H=0.1,~v_S=2~{\rm TeV},~g_{X}=0.6,
\nonumber\\&&~M^2_{Qii}=M^2_Q=1.1~{\rm TeV}^2,~T_{uii}=T_u=1~{\rm TeV},~T_{dii}=T_d=1~{\rm TeV},~(i=1,2,3).
\end{eqnarray}

With the parameters $\tan{\beta}=30$,$~M^2_D=6 ~{\rm TeV}^2$, we plot$~g_{YX}$ versus $\triangle m_{B}$ in the Fig.2(a).$~g_{YX}$ is the coupling constant of gauge mixing that influences the strength of coupling vertexes, and it is the parameter beyond MSSM. In Fig.2(a), the dashed curve corresponds to $M_1=0.8~{\rm TeV}$ and the solid line corresponds to $M_1=1.2~{\rm TeV}$. It can be observed that there is a distinct upward trend in both lines as $g_{YX}$ increases within the range of 0.05 to 0.4, and the growth trend becomes stronger and stronger. The growth of the dashed curve is greater than that of the solid curve.

Supposing $~M^2_D=6 ~{\rm TeV}^2$,$~g_{YX}=0.2$, Fig.2(b) displays a plot of $\triangle m_{B}$ as a function of $M_1$. The solid line and dashed line correspond to $\tan{\beta}=30$ and $\tan{\beta}=20$ in the right diagram. $M_1$ is the mass of the $U(1)_Y$ gaugino, and it gives effect to the mass matrix of neutralino. In Fig.2(b), the two lines are all decreasing functions as $M_1$
turns large in the range of $1~{\rm TeV} <M_1< 2.2~{\rm TeV}$. The dashed line varies from 1.0 to 0.35 and the solid line varies from 1.2 to 0.4.

In summary, it indicates that $M_1$ and $~g_{YX}$ are sensitive parameters for the process of $B^{0}-\bar{B^{0}}$ mixing. When $~g_{YX}$ increases, the values of $\triangle m_{B}$ increase. As $M_1$ decreases, the values of $\triangle m_{B}$ also decrease. Thus, the maximal effects can be achieved when $M_1$ is small and $~g_{YX}$ is large, leading to this conclusion.

\begin{figure}[ht]
\setlength{\unitlength}{5mm}
\centering
\includegraphics[width=3.1in]{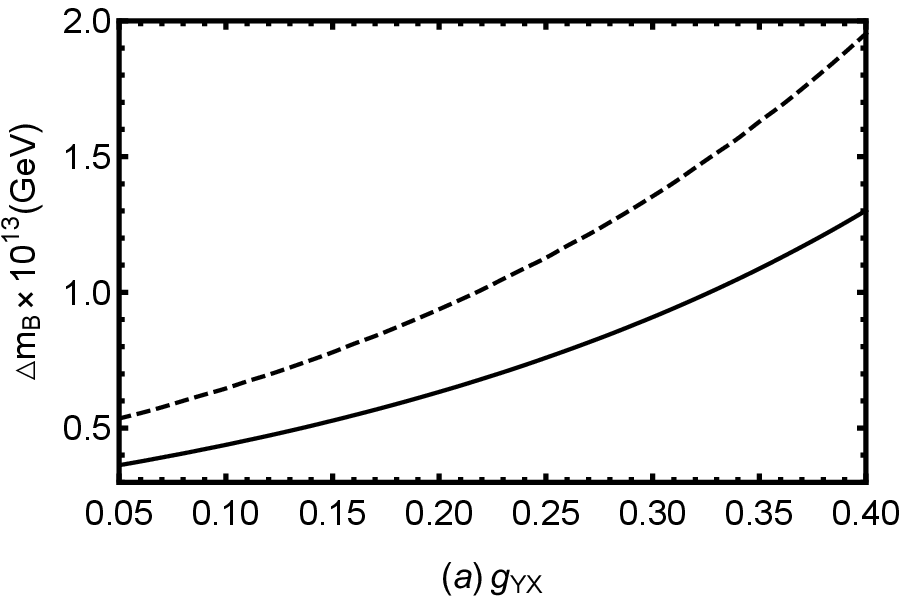}
\vspace{0.2cm}
\setlength{\unitlength}{5mm}
\centering
\includegraphics[width=3.1in]{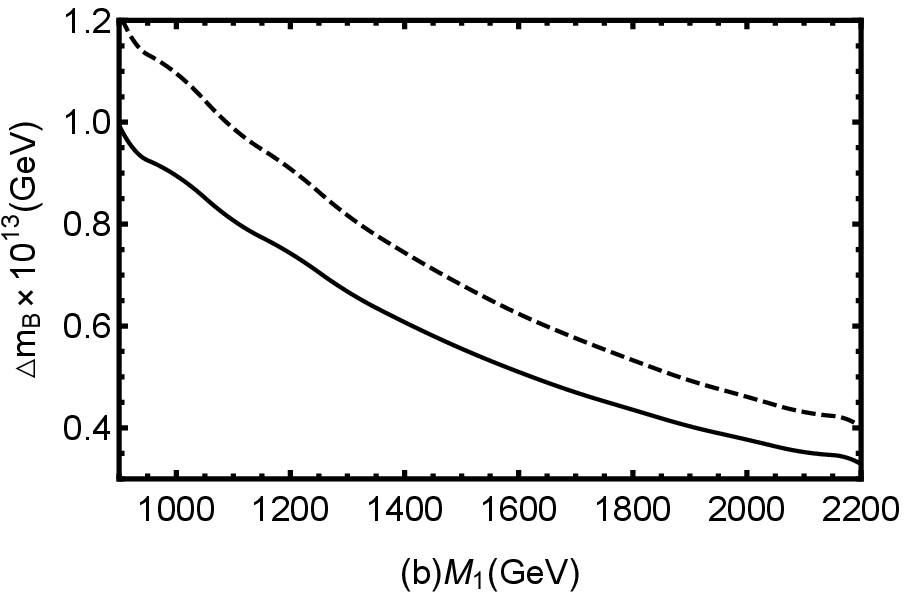}
\caption{(a) displays the solid line ($M_1=1.2~{\rm TeV}$) and dashed line ($M_1=0.8~{\rm TeV}$) in ($g_{YX},\triangle m_{B}$) plane. (b) displays the solid line ($\tan{\beta}=30$) and dashed line ($\tan{\beta}=20$) in ($M_1,\triangle m_{B}$) plane.}{\label {1}}
\end{figure}

We use the parameters as $\tan{\beta}=30$, $~g_{YX}=0.2$,$~M^2_{22}=M^2_{33}=6 ~{\rm TeV}^2$ in Fig.3(a). The solid line ($M_1=1.2~{\rm TeV}$) and dashed line ($M_1=0.8~{\rm TeV}$) represent the relationship between $M^2_{D11}$ and $\triangle m_{B}$ in Fig.3(a). $\triangle m_{B}$ sharply decreases in the range of $3.5 ~{\rm TeV}^2 <M^2_{D11}< 6 ~{\rm TeV}^2$, while rapidly increases in $6 ~{\rm TeV}^2 <M^2_{D11}< 10 ~{\rm TeV}^2$. In the Fig.3(a), the trend of the dashed and solid lines is basically consistent. The gray area is the experimental limit that this process satisfies.

In Fig.3(b), $M^2_{D}$ is set to 6 ${\rm TeV}^2$ and $M_1$ is set to 0.8 ${\rm TeV}$. The solid line ($~g_{YX}=0.3$) and dashed line ($~g_{YX}=0.2$) represent the relationship between $\tan{\beta}$ and $\triangle m_{B}$. It is obvious that both lines have the same tendency to increase and then decrease. When $\tan{\beta}$ is less than 28,
$\triangle m_{B}$ increases as $\tan{\beta}$ increases. However, when $\tan{\beta} > 28$, the situation is just the opposite. The solid curve is larger than the dashed curve. The solid line can reach $1.36\times {10^{ - 13}}~{\rm GeV}$, and the dashed line can almost reach $0.94\times {10^{ - 13}}~{\rm GeV}$.

\begin{figure}[ht]
\setlength{\unitlength}{5mm}
\centering
\includegraphics[width=3.25in]{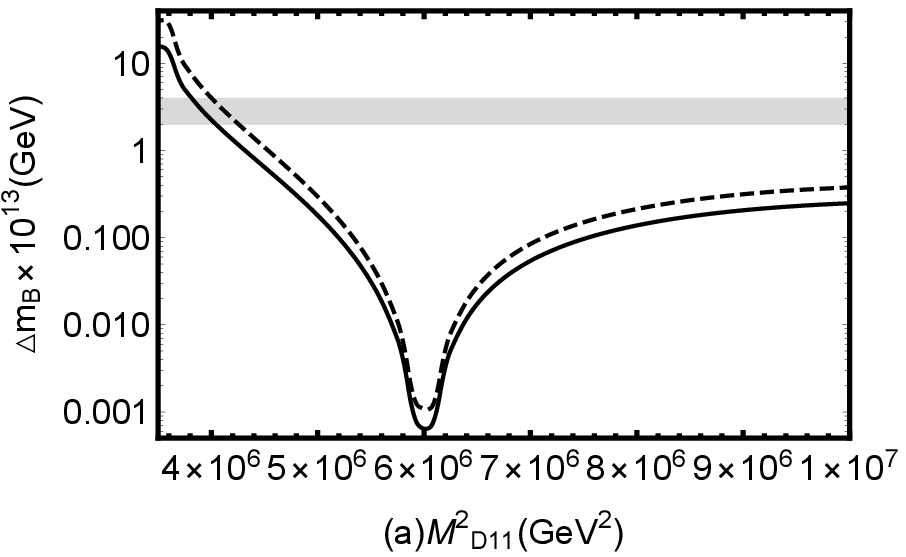}
\vspace{0.2cm}
\setlength{\unitlength}{5mm}
\centering
\includegraphics[width=3.0in]{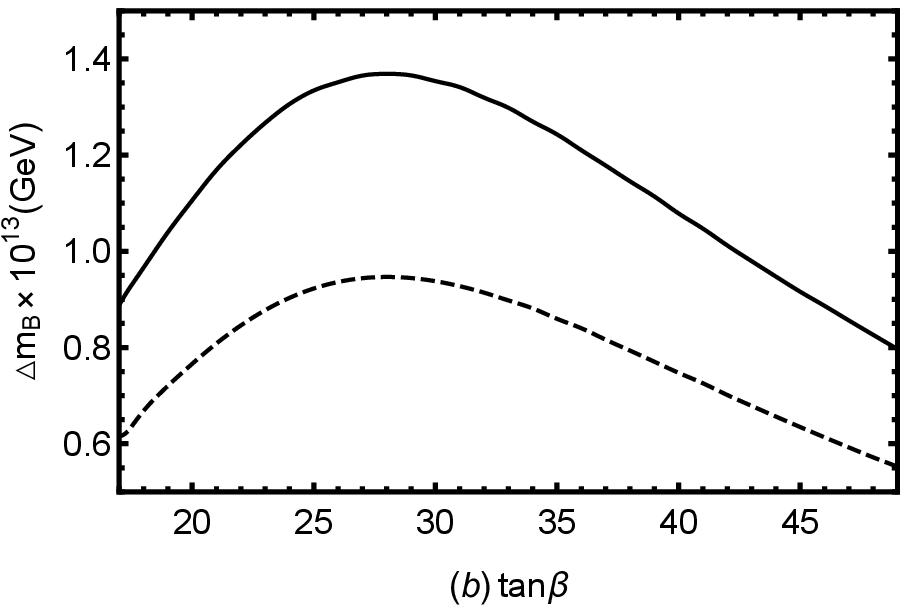}
\caption{(a) displays the solid line ($M_1=1.2~{\rm TeV}$) and dashed line ($M_1=0.8~{\rm TeV}$) in ($M^2_{D11},\triangle m_{B}$) plane. (b) displays the solid line ($~g_{YX}=0.3$) and dashed line ($~g_{YX}=0.2$) in ($\tan{\beta},\triangle m_{B}$) plane.}{\label {2}}
\end{figure}

Based on $g_{YX}=0.2$,$~\tan{\beta}=30$, Fig.4 displays a plot of $\triangle m_{B}$ as a function of $M^2_{Dii}$. The grey area is the experimental limit satisfied by the process. Within the range of $4.5~{\rm TeV}^2<M^2_{Dii}<10~{\rm TeV}^2$, both the dashed line ($\tan{\beta}=20$)and solid line ($\tan{\beta}=30$) show a downward trend, and after being greater than $8~{\rm TeV}^2$, the trend slows down and the two lines gradually overlap.$~M^2_{Dii}$ has an influence on the masses of down-type scalar quark. From the Fig.4, it can be seen that $M^2_{Dii}$ is a sensitive parameter with a significant impact on $\triangle m_{B}$.

\begin{figure}[ht]
\setlength{\unitlength}{5mm}
\centering
\includegraphics[width=3.25in]{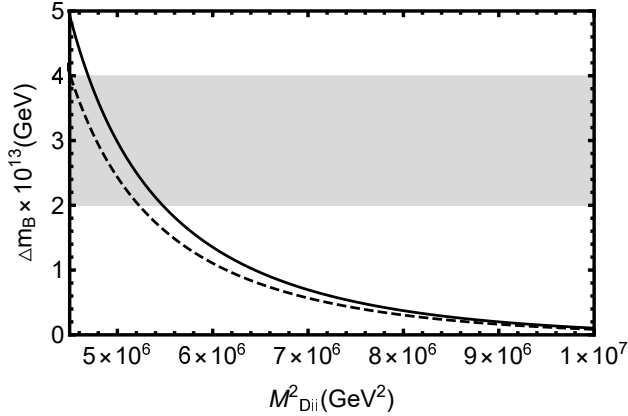}
\vspace{0.1cm}
\caption{The solid line ($\tan{\beta}=30$) and dashed line ($\tan{\beta}=20$) in ($M^2_{Dii},\triangle m_{B}$) plane.}{\label {3}}
\end{figure}

\subsection{The multidimensional scatter plots}

In this subsection, the meanings of shape styles in all the following scatter plots are shown in Table \ref {t12}.
\begin{table}[ht]
\caption{ The meaning of shape style in Fig.5, Fig.6, Fig.7 and Fig.8}
\begin{tabular}{|c|c|}
\hline
Shape style&Fig.5, Fig.6, Fig.7 and Fig.8\\
\hline
\textcolor{blue}{$\blacktriangle$} & $\triangle m_{B}<2\times10^{-16}{~\rm GeV}$\\
\hline
\textcolor{Green}{$\blacksquare$} & $2\times10^{-16}{~\rm GeV}\leqslant\triangle m_{B}<2\times10^{-15}{~\rm GeV}$ \\
\hline
\textcolor{Red}{$\bullet$}&$2\times10^{-15}{~\rm GeV}\leqslant\triangle m_{B}<10^{-12}{~\rm GeV}$  \\
\hline
\end{tabular}
\label{t12}
\end{table}

In Fig.5, the following parameters are used: $\lambda_C=-0.3,~M_2=2.8~{\rm TeV},~{\mu}=1.1~{\rm TeV},~\lambda_H=0.1,~M_1=0.8{~\rm TeV},~M^2_{Dii}=M^2_D=6{~\rm TeV}^2,~v_S=2{~\rm TeV},~(i=1,2,3)$. We randomly scan some parameters, whose ranges are set as :$~0.3<g_{X}<0.8,~2{~\rm TeV}^2<M^2_{Qii}<10{~\rm TeV}^2, ~5<\tan{\beta}<50,~0.01<g_{YX}<0.5,~-4{~\rm TeV}<T_{uii}<4{~\rm TeV},~-4{~\rm TeV}<T_{dii}<4{~\rm TeV}$. From the graph, it can be seen that the \textcolor{Red}{$\bullet$}, \textcolor{Green}{$\blacksquare$}, and \textcolor{blue}{$\blacktriangle$} all cover the entire graph, proving that the parameters $~g_{X}, ~M^2_{Qii},~\tan{\beta},~g_{YX},~T_{uii}$ and $~T_{dii}$ are insensitive and have no significant impact on $\triangle m_{B}$.

\begin{figure}[ht]
\setlength{\unitlength}{5mm}
\centering
\includegraphics[width=3in]{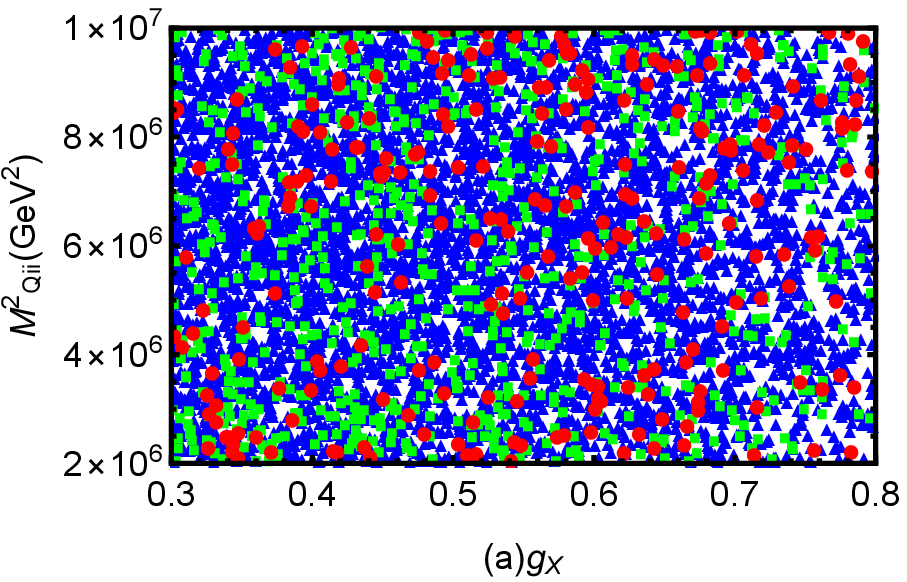}
\vspace{0.2cm}
\setlength{\unitlength}{5mm}
\centering
\includegraphics[width=2.8in]{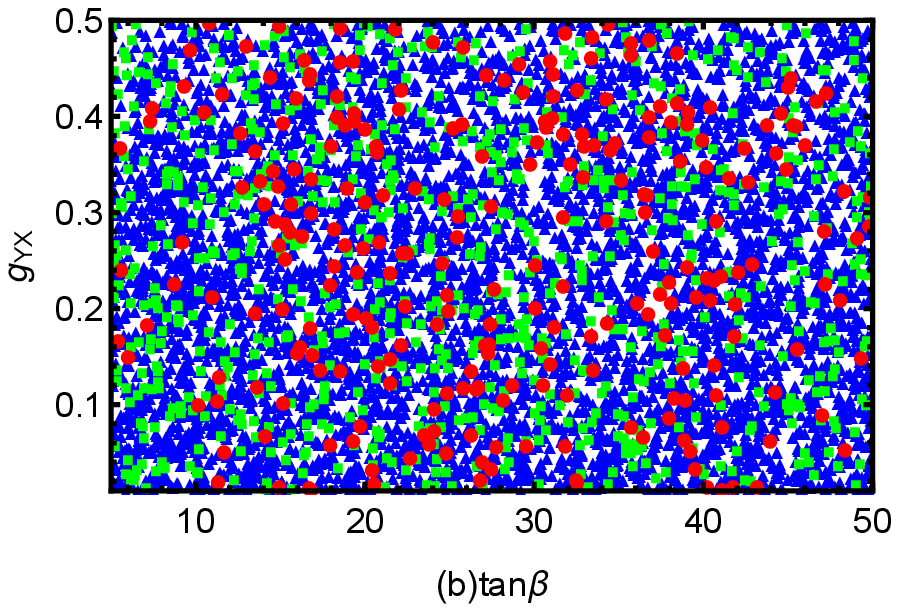}
\setlength{\unitlength}{5mm}
\centering
\includegraphics[width=3in]{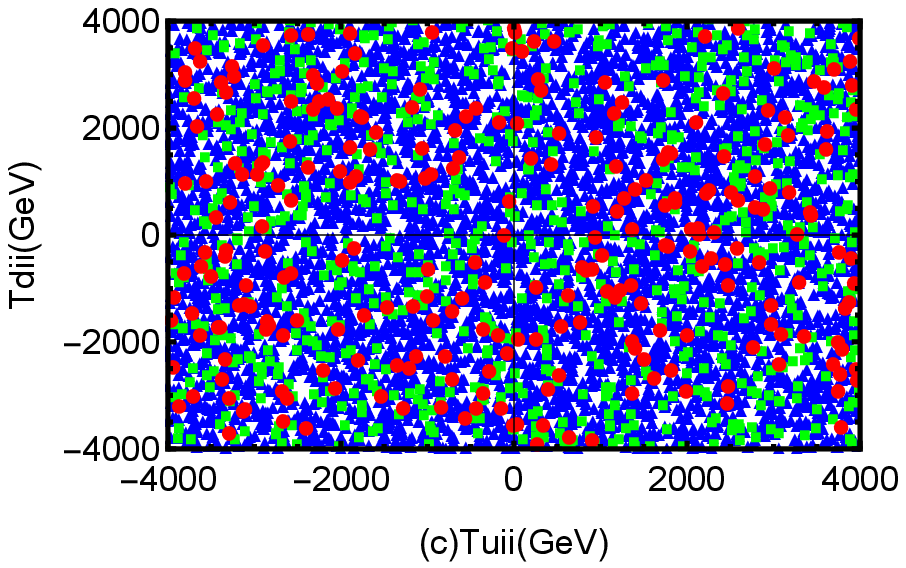}
\caption{$\triangle m_{B}$ in $g_{X}-M^2_{Qii}$ plane(a), $\tan{\beta}-g_{YX}$ plane(b) and $T_{uii}-T_{dii}$ plane(c).}
{\label {5}}
\end{figure}

\begin{figure}[h]
\setlength{\unitlength}{5mm}
\centering
\includegraphics[width=3in]{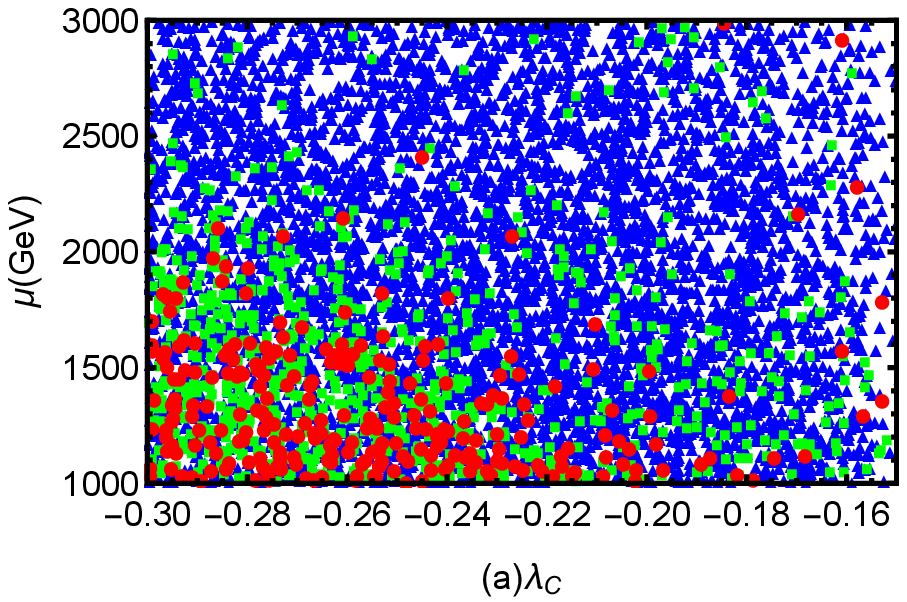}
\vspace{0.2cm}
\setlength{\unitlength}{5mm}
\centering
\includegraphics[width=3in]{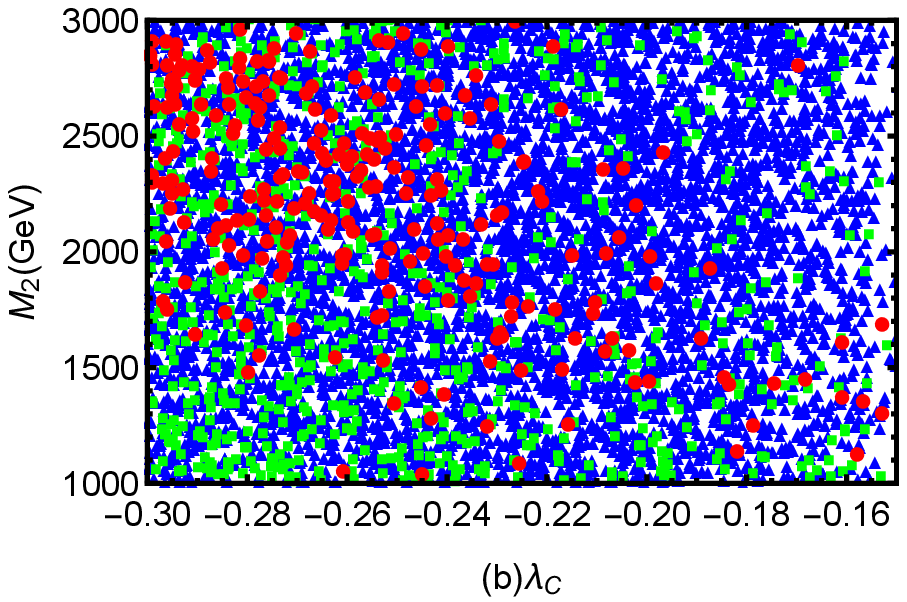}
\setlength{\unitlength}{5mm}
\centering
\includegraphics[width=3in]{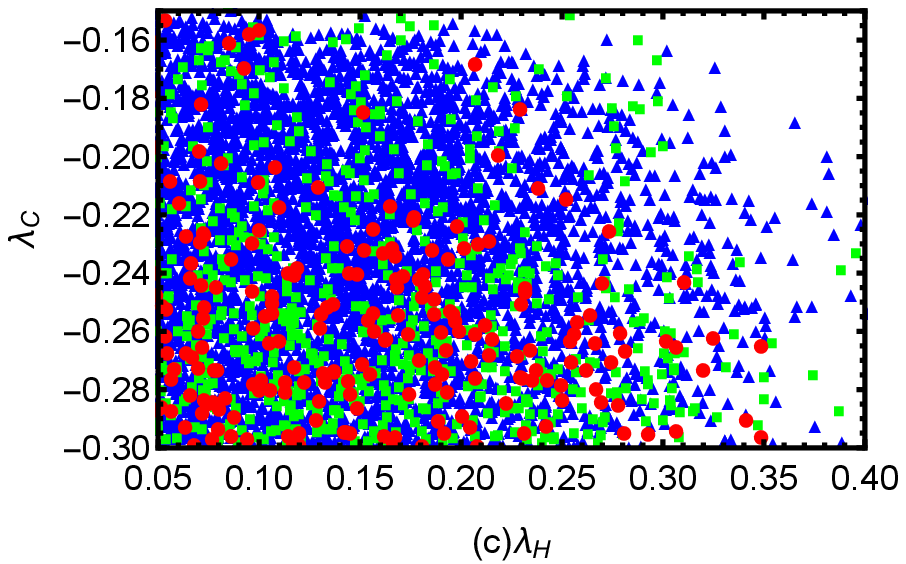}
\caption{$\triangle m_{B}$ in $\lambda_C-{\mu}$ plane(a), $\lambda_C-M_2$ plane(b) and $\lambda_C-\lambda_H$ plane(c).}{\label {4}}
\end{figure}

\begin{figure}[ht]
\setlength{\unitlength}{5mm}
\centering
\includegraphics[width=2.85in]{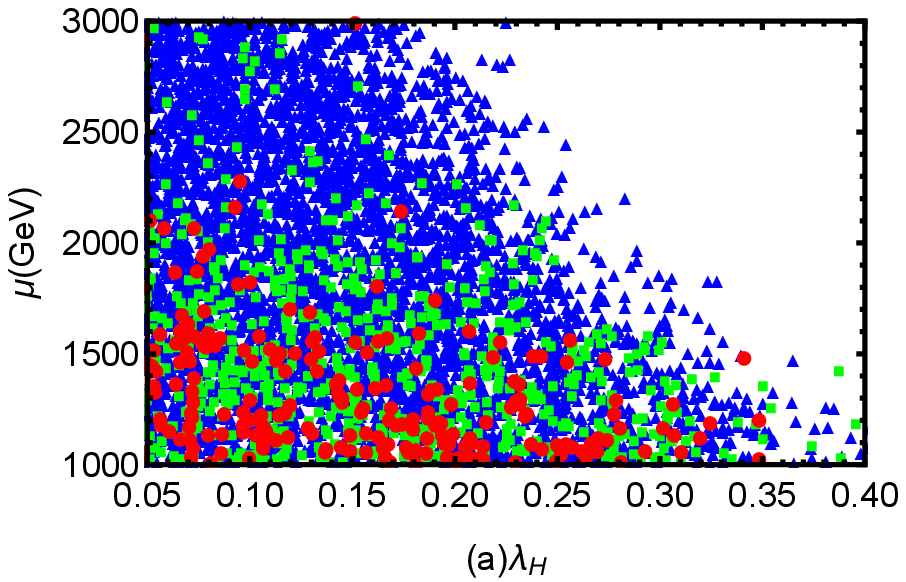}
\vspace{0.2cm}
\setlength{\unitlength}{5mm}
\centering
\includegraphics[width=3in]{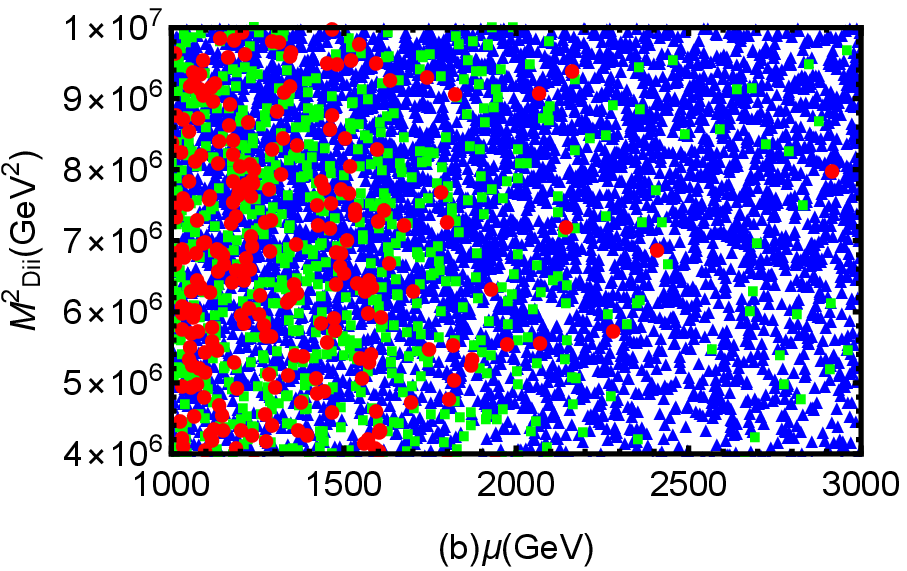}
\setlength{\unitlength}{5mm}
\centering
\includegraphics[width=3in]{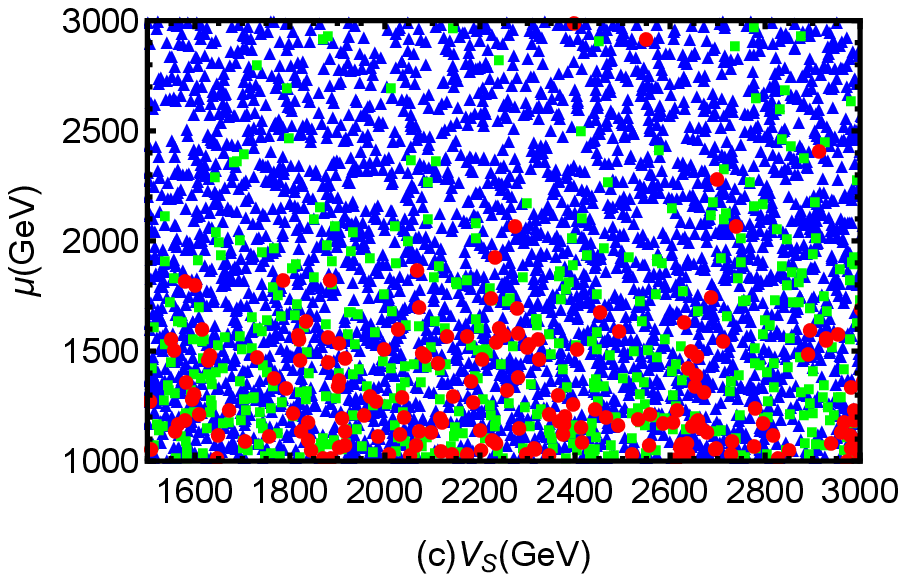}
\caption{$\triangle m_{B}$ in $\lambda_H-{\mu}$ plane(a), ${\mu}-M^2_{Dii}$ plane(b) and ${\mu}-v_S$ plane(c).}
{\label {6}}
\end{figure}

In Fig.6, the set of parameters we will consider are: $M_1=0.8{~\rm TeV}$, $M^2_{Dii}=M^2_D=6{~\rm TeV}^2$, $v_S=2{~\rm TeV}$, $g_{X}=0.6$, $M^2_{Qii}=M^2_Q=6{~\rm TeV}^2$, $\tan{\beta}=30$, $g_{YX}=0.2$, $T_{uii}=T_u=1{~\rm TeV}$, $T_{dii}=T_d=1{~\rm TeV}$, where $i=1,2,3$.

The parameters are randomly scanned over the following ranges: $-0.30<\lambda_C<0.16$, $1{~\rm TeV}<M_2<3{~\rm TeV}$, $1{~\rm TeV}<{\mu}<3{~\rm TeV}$, and $0.05<\lambda_H<0.4$. The meaning of \textcolor{blue}{$\blacktriangle$},~\textcolor{Green}{$\blacksquare$}~and~\textcolor{Red}{$\bullet$} are given in Table \ref {t12}. Supposing $~M_2=2.8{~\rm TeV}$,~$\lambda_H=0.1$, Fig.6(a) show a plot of $\triangle m_{B}$ in the $\lambda_C$ versus ${\mu}$ plane. The space is roughly divided into four parts, the \textcolor{Red}{$\bullet$} is basically located in the bottom left corner of the graph, while the \textcolor{Green}{$\blacksquare$} is mostly located in the bottom left corner, with a small portion in the bottom right corner, except for being covered by the \textcolor{blue}{$\blacktriangle$}. As $\lambda_C$ and ${\mu}$ increase, $\triangle m_{B}$ gradually decreases.

In Fig.6(b), we consider the following parameters: $\lambda_H=0.1$ and ${\mu}=1.1{~\rm TeV}$. We obtain $\triangle m_{B}$ in the plane of the relationship between $\lambda_C$ and $M_2$. The \textcolor{Red}{$\bullet$} is concentrated in areas $2~{\rm TeV} <M_2< 3~{\rm TeV}$ and $-0.3 <\lambda_C< -0.23$. \textcolor{Green}{$\blacksquare$} is mostly located in the $-0.3 <\lambda_C< -0.23$ region, while other parts are scattered. \textcolor{blue}{$\blacktriangle$} covers the entire image. As $\lambda_C$ decreases and $M_2$ increases, $\triangle m_{B}$ gradually increases.

With ${\mu}=1.1{~\rm TeV}$,~$M_2=2.8{~\rm TeV}$, Fig.6(c) depicts an analysis of the effects of parameters $\lambda_C$ and $\lambda_H$. The \textcolor{Red}{$\bullet$} is distributed in $0.05 <\lambda_H< 0.25$ and $-0.3 <\lambda_C< -0.23$. The \textcolor{Green}{$\blacksquare$} is distributed in $-0.3 <\lambda_C< -0.23$, with a small amount distributed above $\lambda_C=-0.23$. The \textcolor{blue}{$\blacktriangle$} is distributed in $0.05 <\lambda_H<0.25$, and a small amount distributed above $\lambda_H=0.25$. As $\lambda_H$ increases, $\triangle m_{B}$ decreases.

The parameters used in Fig.7 are as follows: $M_1=0.8{~\rm TeV}$, $\lambda_C=-0.3$, $M_2=2.8{~\rm TeV}$, $g_{X}=0.6$, $M^2_{Qii}=M^2_Q=6{~\rm TeV}^2$, $\tan{\beta}=30$, $g_{YX}=0.2$, $T_{uii}=T_u=1{~\rm TeV}$, $T_{dii}=T_d=1{~\rm TeV}$, where $i=1,2,3$.

With $v_S=2{~\rm TeV}$,$~M^2_D=6{~\rm TeV}^2$, Fig.7(a) displays a plot of $\triangle m_{B}$ in the $\lambda_H$ versus ${\mu}$ plane. We can clearly see that the space is roughly divided into four parts, with most of the \textcolor{Red}{$\bullet$} in areas $0.05 <\lambda_H<0.35$ and $1~{\rm TeV} <{\mu}< 1.6~{\rm TeV}$, while the \textcolor{Green}{$\blacksquare$} is basically the same as the \textcolor{Red}{$\bullet$}, but the quantity is more than the \textcolor{Red}{$\bullet$}. The \textcolor{blue}{$\blacktriangle$} occupies the left side, resembling a right angled trapezoid, with no point distribution in the upper right corner. It can be concluded that both $\lambda_H$ and ${\mu}$ are sensitive parameters and act together on $\triangle m_{B}$.

We suppose the parameters with $\lambda_H=0.1$, $v_S=2{~\rm TeV}$ in Fig.7(b). We show $\triangle m_{B}$ in the plane of ${\mu}$ and $M^2_D$. The majority of \textcolor{Red}{$\bullet$} distributes in $1~{\rm TeV} <{\mu}< 1.6~{\rm TeV}$, while the majority of \textcolor{Green}{$\blacksquare$} distributes in $1~{\rm TeV} <{\mu}< 1.9~{\rm TeV}$. The \textcolor{blue}{$\blacktriangle$} covers the entire graph. As ${\mu}$ increases, the number of \textcolor{Red}{$\bullet$} and \textcolor{Green}{$\blacksquare$} decreases, while $\triangle m_{B}$ shows a gradual decreasing trend.

With $\lambda_H=0.1$, $M^2_D=6{~\rm TeV}^2$, Fig.7(c) presents an analysis of the effects of parameters ${\mu}$ and $v_S$. The entire image is divided into three parts, with the \textcolor{Red}{$\bullet$} mostly distributes in $1~{\rm TeV} <{\mu}< 1.5~{\rm TeV}$, the \textcolor{Green}{$\blacksquare$} mostly distributes in $1~{\rm TeV} <{\mu}< 1.9~{\rm TeV}$, and the \textcolor{blue}{$\blacktriangle$} occupies the entire image.

With $M_1=0.8{~\rm TeV},~\lambda_C=-0.3,~v_S=2{~\rm TeV},~\lambda_H=0.1,~g_{X}=0.6,~{\mu}=1.1{~\rm TeV},
~M^2_{Qii}=M^2_Q=6{~\rm TeV}^2,~\tan{\beta}=30,~g_{YX}=0.2,~T_{uii}=T_u=1{~\rm TeV},~T_{dii}=T_d=1{~\rm TeV},(i=1,2,3)$in Fig.8. We show $\triangle m_{B}$ in the plane of the relationship between $M_2$ and $M^2_{Dii}$. The \textcolor{Red}{$\bullet$} is within $1.7~{\rm TeV} <M_2< 3~{\rm TeV}$, \textcolor{Green}{$\blacksquare$} and \textcolor{blue}{$\blacktriangle$} basically cover the entire image.

\begin{figure}[ht]
\setlength{\unitlength}{5mm}
\centering
\includegraphics[width=3in]{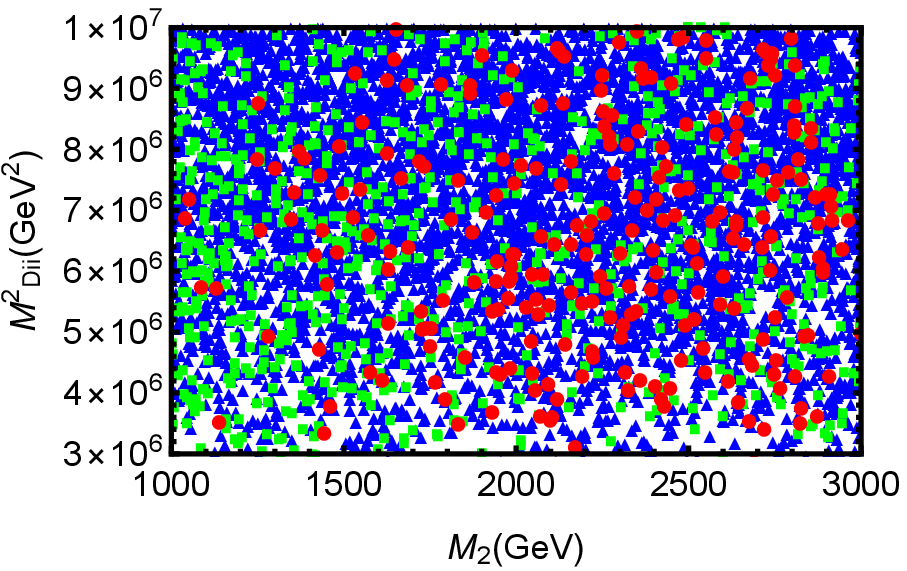}
\vspace{0.2cm}
\caption{$\triangle m_{B}$ in $M_2-M^2_{Dii}$ plane.}
{\label {7}}
\end{figure}

By analying, the above graphs, it is evident that the parameters $v_S$, $M^2_D$, $\lambda_C$, ${\mu}$, $M_2$ and $\lambda_H$ are relatively sensitive for the process of $B^{0}-\bar{B^{0}}$ mixing. When $M_2$ decreases and $~{\mu},~\lambda_C,~\lambda_H$ increases, $\triangle m_{B}$ can obtain better results.

\section{Conclusion}
In this paper, we apply the basic formulas and research methods of neutral meson mixing to study the mixing of $B^{0}-\bar{B^{0}}$  in the $U(1)_X$SSM. Using the effective Hamiltonian method, the Wilson coefficients and mass difference $\triangle m_{B}$ are derived. We obtain abundant numerical results by scanning parameter space when taking into account the latest experimental limitations.

In numerical calculation, we obtain rich data by scanning large parameter spaces. After processing the data, we obtain interesting one-dimensional and multidimensional scatter plots. We select the parameters $~\lambda_C,~M_1,~M_2,~{\mu},~\lambda_H,~M^2_{Djj}=M^2_D,~M^2_{D11},~v_S,~g_{X},~M^2_{Qii}=M^2_Q
,~\tan{\beta},~g_{YX},~T_{uii}=T_u,~T_{dii}=T_d,(i=1,2,3),(j=2,3)$ as variables. Through the analysis of numerical results, we find that $~v_S,~M^2_D,~\lambda_C,~{\mu},~M_2,~\tan{\beta},~g_{YX},~M_1$ and $~\lambda_H$ are sensitive parameters. Within the parameter range specified in this paper,$~\triangle m_{B}$ is an decreasing function of $~M_2,~M^2_D,~M_1,~{\mu}$.$~\triangle m_{B}$ is an increasing function of $~\lambda_C,~\lambda_H$ and $g_{YX}$. The parameters $~g_{X}, ~M^2_{Qii},~\tan{\beta},~T_{uii}$ and $~T_{dii}$ are insensitive and have no significant impact on $\triangle m_{B}$.
The theoretical predicted value of mass splitting $\triangle m_{B}$ in a specific parameter space can be in good agreement with experimental results, providing new ideas for searching for NP.

\begin{acknowledgments}

This work is supported by National Natural Science Foundation of China (NNSFC)
(No.12075074), Natural Science Foundation of Hebei Province
(A2020201002, A202201022, A2022201017), Natural Science Foundation of Hebei Education Department (QN2022173), Post-graduate's Innovation Fund Project of Hebei University (HBU2023ss043), the youth top-notch talent support program of the Hebei Province.
\end{acknowledgments}

\appendix
\section{Used coupling in $U(1)_X$SSM}
In the basis $ \left(\tilde{W}^-, \tilde{H}_d^-\right), \left(\tilde{W}^+, \tilde{H}_u^+\right)$, the definition of the mass matrix for chargino is given by
\begin{equation}
M_{\tilde{\chi}^\pm} = \left(
\begin{array}{cc}
M_2 &\frac{1}{\sqrt{2}} g_2 v_u \\
\frac{1}{\sqrt{2}} g_2 v_d  &\frac{1}{\sqrt{2}} {\lambda}_{H} v_S  + \mu\end{array}
\right).
\label{mxzf}
\end{equation}

In the basis $\left(\tilde{d}^0_{L,{{\alpha_1}}}, \tilde{d}^0_{R,{{\alpha_2}}}\right), \left(\tilde{d}^{0,*}_{L,{{\beta_1}}}, \tilde{d}^{0,*}_{R,{{\beta_2}}}\right)$, the definition of the squared mass matrix for down type squark is given by
\begin{equation}
M^2_{\tilde{D}} = \left(
\begin{array}{cc}
m_{\tilde{d}_L^0\tilde{d}_L^{0,*}} &m^\dagger_{\tilde{d}_R^0\tilde{d}_L^{0,*}}\\
m_{\tilde{d}_L^0\tilde{d}_R^{0,*}} &m_{\tilde{d}_R^0\tilde{d}_R^{0,*}}\end{array}
\right),
\end{equation}
where
\begin{eqnarray}
&&m_{\tilde{d}_L^0\tilde{d}_L^{0,*}} = \frac{1}{24}
 \Big( (3 g_{2}^{2}  + g_{1}^{2} + g_{Y X}^{2}) ( v_{u}^{2}- v_{d}^{2}  ) +  g_{Y X} g_{X}  (2 v_{\bar{\eta}}^{2}  -2 v_{\eta}^{2}  - v_{d}^{2}  + v_{u}^{2})\Big)+m_{\tilde{Q}}^2  +\frac{ v_{d}^{2}}{2} {Y_{d}^{\dagger}  Y_d},\nonumber\\
&&m_{\tilde{d}_L^0\tilde{d}_R^{0,*}} = -\frac{1}{2}  \Big(\sqrt{2}  (- v_d T_d  + v_u Y_d \mu^* ) + v_u v_S Y_d {\lambda}_{H}^* \Big),\nonumber\\
&&m_{\tilde{d}_R^0\tilde{d}_R^{0,*}} = \frac{1}{24}   \Big(2  (g_{1}^{2} + g_{Y X}^{2}) ( v_{u}^{2}  - v_{d}^{2})+ g_{Y X} g_{X} (4 v_{\bar{\eta}}^{2}  - 4 v_{\eta}^{2}  +5 v_{u}^{2}  - 5 v_{d}^{2} ),\nonumber \\
&&\hspace{1.8cm}+3 g_{X}^{2} (2 v_{\bar{\eta}}^{2}  - 2 v_{\eta}^{2}  + v_{u}^{2}  - v_{d}^{2})\Big) +m_{\tilde{D}}^2  +\frac{ v_{d}^{2}}{2} {Y_{d}^{\dagger}  Y_d}.\nonumber
\end{eqnarray}

In the basis $\left(\tilde{u}^0_{L,{{\alpha_1}}}, \tilde{u}^0_{R,{{\alpha_2}}}\right), \left(\tilde{u}^{0,*}_{L,{{\beta_1}}}, \tilde{u}^{0,*}_{R,{{\beta_2}}}\right)$, the definition of the squared mass matrix for up type squark is
\begin{equation}
M^2_{\tilde{U}} = \left(
\begin{array}{cc}
m_{\tilde{u}_L^0\tilde{u}_L^{0,*}} &m^\dagger_{\tilde{u}_R^0\tilde{u}_L^{0,*}}\\
m_{\tilde{u}_L^0\tilde{u}_R^{0,*}} &m_{\tilde{u}_R^0\tilde{u}_R^{0,*}}\end{array}
\right),
\end{equation}
where
\begin{eqnarray}
	&&m_{\tilde{u}_L^0\tilde{u}_L^{0,*}} = \frac{1}{24}  \Big( (g_{1}^{2} -3 g_{2}^{2}+ g_{Y X}^{2}) ( v_{u}^{2}- v_{d}^{2}  ) +   g_{Y X} g_{X}  (2 v_{\bar{\eta}}^{2}  -2 v_{\eta}^{2}  - v_{d}^{2}  + v_{u}^{2})\Big)
+  m_{\tilde{Q}}^2  +\frac{ v_{u}^{2}}{2} {Y_{u}^{\dagger}  Y_u},\nonumber\\
	&&m_{\tilde{u}_L^0\tilde{u}_R^{0,*}} = -\frac{1}{2}   \Big(\sqrt{2}  (v_d Y_u \mu^*  - v_u T_u ) + v_d v_S Y_u {\lambda}_{H}^* \Big),\nonumber\\
	&&m_{\tilde{u}_R^0\tilde{u}_R^{0,*}} = \frac{1}{24}   \Big(4  (g_{1}^{2} + g_{Y X}^{2}) (- v_{u}^{2}  + v_{d}^{2})+  g_{Y X} g_{X}  (7 v_{d}^{2}  -7 v_{u}^{2}  -8 v_{\bar{\eta}}^{2}  + 8 v_{\eta}^{2} )\nonumber \\
	&&\hspace{1.8cm}+3  g_{X}^{2} (-2 v_{\bar{\eta}}^{2}  + 2 v_{\eta}^{2}  - v_{u}^{2}  + v_{d}^{2})\Big) +  m_{\tilde{U}}^2  +\frac{ v_{u}^{2}}{2} {Y_{u}^{\dagger}  Y_u}.\nonumber
\end{eqnarray}

Here, we show the needed couplings in this work.
\begin{eqnarray}
&&\mathcal{L}_{d\tilde{u}\chi^{-}}=d_{i}\tilde{u}_{k}^*
\Big(Y_{d,i}Z_{k,i}^{U,*}U_{j2}^{*}P_L+(V_{j2}Y_{u,i}^{*}Z_{k,3+i}^{U,*}-g_2V_{j1}Z_{k,i}^{U,*}
)P_R\Big)\bar{\chi}_j^{-},
\\&&\mathcal{L}_{\tilde{d}d\chi^0}=-\frac{1}{6}d_i\tilde{d}^*_k\Big\{\Big[ (2\sqrt{2}g_1) Z_{k,3+i}^{D,*} N_{j1}^{*}
 + \sqrt{2}(2 g_{Y X}  + 3 g_{X})Z_{k,3+i}^{D,*}N_{j5}^{*}+6 Y_{d,i}  Z_{k,i}^{D,*}  N_{j3}^{*}\Big]P_L\nonumber\\
&&\hspace{1.5cm}+\Big[6 N_{j3} Y_{d,i}^{*} Z_{k,3+j}^{D,*}+ \sqrt{2}Z_{k,i}^{D,*}( g_1 N_{j1}
-3 g_2 N_{j2}    + g_{Y X} N_{j5})   \Big]P_R\Big\}\bar{\chi}^0_j.
\end{eqnarray}

\end{document}